\documentclass[prl,twocolumn,showpacs,floatfix]{revtex4}
\usepackage{bm}
\usepackage{amssymb}
\begin{document}
\title{
Orbital evolution of a test particle around a black hole: higher-order
corrections}
\author{Lior M.\ Burko}
\affiliation{Department of Physics, University of Utah, Salt Lake City,
Utah 84112}
\date{draft, August 13, 2002}
\begin{abstract}

We study the orbital evolution of a radiation-damped binary in the
extreme mass ratio limit, and the resulting waveforms, to one order beyond
what can be obtained using the conservation laws approach. The equations
of motion are solved perturbatively in the mass ratio (or the corresponding
parameter in the scalar field toy model), using the self force, for
quasi-circular  orbits around a Schwarzschild black hole. This approach is
applied for the scalar model. Higher-order corrections yield a phase shift 
which, if included, may make gravitational-wave astronomy potentially highly
accurate.
\end{abstract}
\maketitle

The problem of determining the orbital evolution of a binary which
undergoes radiation damping has become timely and crucial, as earth-based
gravitational-wave detectors (LIGO/VIRGO) will become operative in the
near future, and as a space-based detector (LISA) is currently planned to
fly in about a decade \cite{lisa}. The orbital evolution becomes a much
more feasible problem when $M\gg \mu$, $M$ being the mass of the central
black hole, and $\mu$ that of the inspiralling companion. For
typical parameters of a supermassive black hole at
galactical centers and a stellar black hole, with $M/\mu\approx 10^{5-7}$, 
the frequency of the emitted waves is in the good-sensitivity band for
space-based detectors. LISA is expected to measure such waves with
amplitude signal-to-noise ratio of order $10-100$ \cite{finn-thorne} with
one year integration time, with occurrence rate of a few per year to a few
per month \cite{sigurdsson-rees}. 

For $M\gg \mu$, the companion can be modeled as 
a pointlike particle, and the orbital evolution can be studied
using perturbation theory: $\mu$ moves in the fixed
background spacetime of $M$; because of the
(small) perturbation to the metric due to $\mu$, the latter does not
move along a geodesic of the background. Instead, it moves along an
accelerated orbit, resulting 
from its self force (SF). (In an alternative equivalent picture, 
the particle moves along a geodesic of a perturbed spacetime.) 

The calculation of the SF for realistic astrophysical
systems turns out to be rather difficult. Because of that difficulty, it
is generally hoped that simpler methods, applicable for simplified
systems, may be useful. Specifically, an approach dubbed
``radiation reaction (RR) without radiation reaction forces''
(RRWORRF) has been used to evolve the orbit and obtain the corresponding
waveforms  \cite{hughes,kennefick}. The underlying motivation for that
approach is roughly the following: So long as the orbital evolution is
adiabatic, the constants of motion (COM) which describe the orbit change
only slowly. The rate of change of the COM 
can be inferred from the flux of the corresponding quantities to
infinity and down the event horizon of the central black hole, and from
this rate of change one can infer the orbital evolution. One complication
to this strategy is that one of the COM, the Carter constant $Q$, is
non-additive. It is generally impossible to
use balance arguments to find $\dot{Q}$. For that reason, this
approach has been used for cases where $\dot{Q}$ is trivial, such as for
circular \cite{hughes} or equatorial \cite{kennefick} orbits around a Kerr
black hole. The conventional approach is to 
compute the local SF acting on $\mu$, $f_{\alpha}^{\rm SF}$, and
infer from it $\dot Q$. With the knowledge of the rates of change of all
COM, one would be able to evolve the orbit and find the
corresponding waveform. That approach, however, can only find the orbital
evolution to leading order in $\mu/M$. [In what follows, we denote
$\epsilon=\mu/M$. For
the case in which the particle carries a scalar or electric charge $q$,
$\epsilon=q^2/(\mu M)$.] Consequently, conservation laws (and 
conservation of $\mu$) are violated to
$O(\epsilon^2)$, and meaningful quantities such as the number of cycles
${\cal N}_{\rm cyc}$ 
spent in a logarithmic interval of frequency $f$,  
$\,d{\cal N}_{\rm cyc}/ \,d\ln f$,  and the orbit are found to
$O(\epsilon^{-1})$. 
It can be expected that corrections of $O(1)$ to $\,d{\cal N}_{\rm
cyc}/\,d\ln f$ and the orbit may be of practical importance: when the data
stream is cross correlated with a theoretical template, the cross
correlation plummets when the two slip by too much. Already when the two
lose phase by as little as one half-cycle, their overlap integral will be
strongly reduced \cite{cutler-93}. High accuracy
theoretically-derived templates are not necessary for detection of the
waves. They are needed, nevertheless, for precision gravitational-wave
astronomy. (See also \cite{burko-ijmpa}.) It is also important to provide
estimates for the error in the first-order waveforms. 

It is hard to generalize the RRWORRF program to handle
higher-order corrections. The reason being that the RRWORRF program, and
indeed any nonlocal approach based on conservation laws, ignores
conservative SF effects, because the fields associated with
the latter decay faster at infinity than those associated with
dissipation. When integrating over a sphere at infinity, the conservative
effects are in practice discarded. 

In this paper we propose to use the SF directly to find the
orbit to $O(1)$. Dimensional-analysis arguments suggest that corrections to
the orbit-integrated $\,d{\cal N}_{\rm cyc}/\,d\ln f$ are at order unity. We
present a detailed analysis of a simplified toy model, which allows us to
calculate this correction accurately. We find that the naive analysis
overestimates the effect of interest by a full order of magnitude.

For simplicity, we apply the approach for quasi-circular equatorial orbits
around a Schwarzschild black hole. The metric in the usual Schwarzschild
coordinates is given by
\begin{equation}
\,ds^2=-\left(1-\frac{2M}{r}\right)\,dt^2+\left(1-\frac{2M}{r}\right)^{-1}\,dr^2
+r^2\,d\Omega^2\, .
\label{metric}
\end{equation}
Here, $\,d\Omega^2=\, d\vartheta^2 +\sin^2\vartheta\,d\phi^2$. 
Our approach is based on finding $\dot{r}(t)$ and  $\omega (t)$
perturbatively to $O(\epsilon^2)$, $\omega$ being the orbital 
frequency. Henceforth, we denote by an overdot and
a prime (partial) differentiation with respect to coordinate time $t$ and
$r$, respectively. 

We use the normalization condition for $u^{\alpha}$, 
namely $u^{\alpha}u_{\alpha}=-1$, to eliminate $u^t$ from the equations of
motion (EOM) $\,Du^{i} / 
\,d\tau=\mu^{-1} f_{k}^{\rm SF}g^{ik}$, $i,k=t,r,\phi$, where $D$
denotes covariant differentiation compatible with the metric
(\ref{metric}). We next use the $t$ component of the EOM  to
eliminate ${\dot u}^t$.  
We can simplify the EOM to first-order (nonlinear) ODEs by taking
$\dot{r}=V(r)$, $\dot{x}=Vx'(r)$, and $x$ denotes any quantity.
We find the EOM to be 
\begin{eqnarray}
VV'&-&\frac{3MV^2}{r(r-2M)}-(r-2M)\sigma-
\frac{1}{\mu {u^t}^2}\left[\left(1-\frac{2M}{r}\right)f_r^{\rm SF}
\right. \nonumber \\ 
&+&\left.\frac{V}{1-2M/r}f_t^{\rm SF}\right]=0
\\
V\sigma '&-&\frac{3MV}{r^4}+2\frac{M/r^3+\sigma}{1-2M/r}
\left[\frac{2}{r}V\left(1-\frac{3M}{r}\right)\right. \nonumber \\
&-&\left.\frac{f_t^{\rm SF}}{\mu {u^t}^2}\right]
-\frac{2(M/r^3+\sigma)^{1/2}}{\mu {u^t}^2r^2}
f_{\phi}^{\rm SF}=0\, ,
\end{eqnarray}
where ${u^t}^2=1/[1-3M/r-r^2\sigma-V^2/(1-2M/r)]$. 
Here, $\sigma$ measures the deviation from
Kepler's law, i.e., $\omega^2=M/r^3+\sigma (r)$. This relation
is gauge independent, although each of the terms on the right hand side
are separately gauge dependent \cite{detweiler-private}. We next
expand in powers of $\epsilon$ as $\sigma (r)=\sigma_{(1)}+\sigma_{(2)}$,
$V= V_{(1)}+ V_{(2)}$, and $a_{i}= a^{(1)}_i+ a^{(2)}_i$, $x_{(j)}$
denoting the term in $x$ which is at $O(\epsilon^{j})$, and $a_i$ being
the self acceleration. We then expand the self force as 
$f_{i}^{\rm SF}= f^{(1)}_i+ f^{(2)}_i$, where $f^{(j)}_i=\mu a^{(j)}_i$. 
Solving perturbatively, we find that
\begin{equation}
\sigma_{(1)}=-\frac{r-3M}{\mu r^2}f^{(1)}_{r}
\end{equation}
\begin{equation}
V_{(1)}=\frac{2r}{\mu M}\frac{r-3M}{r-6M}
\left[\left(\frac{M}{r}\right)^{\frac{1}{2}}\left(1-\frac{2M}{r}\right)
f^{(1)}_{\phi}+Mf^{(1)}_{t}\right]
\end{equation}
\begin{eqnarray}
V_{(2)}&=&\frac{r(r-3M)}{\mu^2M^2(r-6M)^2}
\left[2\left(\frac{M}{r}\right)^{\frac{1}{2}}f^{(1)}_{\phi}f^{(1)'}_{r}
r(r-2M)^2\right. \nonumber \\
&\times&
(r-3M)+\left(\frac{M}{r}\right)^{\frac{1}{2}}f^{(1)}_{\phi}f^{(1)}_{r}
(5r-6M)(r-2M) \nonumber \\
&\times& (r-3M)+2Mf^{(1)}_{t}f^{(1)'}_{r}r^2(r-2M)(r-3M) \nonumber \\
&+&4Mf^{(1)}_{t}f^{(1)}_{r}r^2(r-3M) +2\mu M^2f^{(2)}_{t}(r-6M)\nonumber
\\ 
&+& \left.
2\mu \left(\frac{M}{r}\right)^{\frac{3}{2}}f^{(2)}_{\phi}
(r-2M)(r-6M)\right] \, .
\label{v2}
\end{eqnarray}
Notice, that Eq.~(\ref{v2}) is only a formal expression, as we do not know 
$f^{(2)}_{i}$. [Currently, $f^{(1)}_{i}$ are known for scalar field
RR for circular orbits around Schwarzschild \cite{burko-00}. For
gravitational RR, even  $f^{(1)}_{i}$ are not known as yet.] This,
however, is not an important problem: The coefficients of the unknown 
terms in Eq.~(\ref{v2}) are much smaller than the coefficients of the
other terms, such that their relative contribution is small. 
[From dimensional-analysis arguments, it can be shown that   
$f_t^{(2)}\sim \alpha_t(M/\mu){f_t^{(1)}}^2$, 
$f_r^{(2)}\sim \alpha_r(M/\mu){f_r^{(1)}}^2$, and 
$f_{\phi}^{(2)}\sim \alpha_{\phi}(M/r)^{3/2}\mu^{-1}{f_{\phi}^{(1)}}^2$,
and consider their contribution with $\alpha_i$ not much
greater than unity. In deriving these relations recall that it is simplest to
analyze the scaling of the four-accelerations, and only then to obtain the
forces. Notice also that these terms vanish at $r=6M$.] We shall
henceforth happily ignore the terms involving $f^{(2)}_{i}$. 

We can study the importance of the higher-order correction by considering
two dimensionless quantities, $\,d{\cal N}_{\rm cyc}/\,d(\ln f)$ and
$V/(r\omega)$. Then, we compare these quantities between a theoretical
template accurate to $O(\epsilon^{-1})$, and a template accurate to
$O(1)$. [Modeling the actual data stream by the $O(1)$ template, we can
test the $O(\epsilon^{-1})$ template.] 
$V/(r\omega)$ is related to the rate of change of the
envelope of the chirp wave. Notice that $\,d{\cal N}_{\rm
cyc}/\,d(\ln f)\equiv\omega^2/[(2)\pi{\dot\omega}]$ is gauge
independent. The
difference in these quantities 
between the $O(1)$ expressions and their $O(\epsilon^{-1})$ counterparts
can be expanded using the expressions above. We find that 
\begin{equation}
\Delta \frac{\,d{\cal N}_{\rm cyc}}{\,d(\ln f)}=
-\frac{2}{(2)3\pi}\sqrt{\frac{M}{r}}
\left[\frac{3}{2}\frac{r^3\sigma_{(1)}}{MV_{(1)}}+\frac{1}{3}
\frac{r^4\sigma_{(1)}'}{MV_{(1)}}-\frac{V_{(2)}}{V_{(1)}^2}
\right]\, ,
\label{B}
\end{equation}
and
\begin{equation}
\Delta\left(\frac{V}{r\omega}\right) / 
\left(\frac{V}{r\omega}\right)^2=\sqrt{\frac{M}{r}}
\left[\frac{1}{2}\frac{r^3\sigma_{(1)}}{MV_{(1)}}-\frac{V_{(2)}}{V_{(1)}^2}
\right]\, .
\label{A}
\end{equation}  
Notice that the last two quantities are at $O(1)$. 
The bracketed factor of $2$ should be
introduced for scalar field RR (and is absent for gravitational RR). 
Notice also that Eqs. (\ref{B}) and (\ref{A}) do not depend on
$\sigma_{(2)}$. 

The orbit can be obtained by integrating 
\begin{equation}
t(r)=\int_{r_{\rm start}}^{r}\frac{1}{V({\tilde r})}\,d{\tilde r}
\;\;\; {\rm and} \;\;\;
\phi (t)=\int_{r_{\rm start}}^{r}\frac{\omega
({\tilde r})}{V({\tilde r})}\,d{\tilde r}\, . 
\end{equation} 
The
integrands can be expanded to $O(1)$
using the expressions above.  These integrals are easy to do using
fourth-order Runge-Kutta integration. The triad $r,t(r),\phi(r)$ can then
be inverted to yield $t,r(t),\phi(t)$, from which the orbit and the
waveform can be reconstructed. 

\begin{figure}
\input epsf
\epsfxsize=8.5cm
\centerline{
\epsfbox{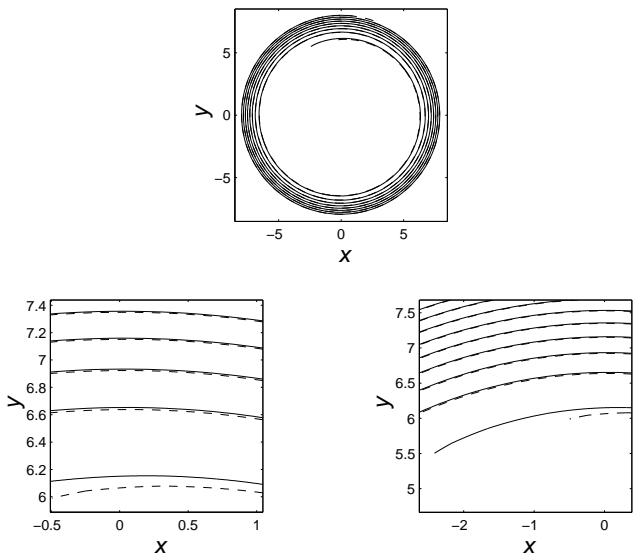}}
\caption{The orbit of a scalar charge with $q^2/(\mu M)=0.1$. The upper
panel shows the last few orbits before the ISCO. The solid (dashed) line
is the orbit to $O(1)$ [$O(\epsilon^{-1})$]. The lower panels display
two enlargements of the same orbits: the one on the left emphasizes the  
difference in the $r$ values, and the one on the right the difference in
phase. Here, $x=(r/M)\cos\phi$ and $y=(r/M)\sin\phi$.}
\label{fig1}
\end{figure}

As currently we do not know $f^{(1)}_i$ for gravitational RR, we
specialize next to scalar field RR. [When $f^{(1)}_i$ are obtained for
gravitational RR in any regular gauge, the analysis below can be repeated
for that case.] Specifically, we study the model of a scalar charge $q$
of mass $\mu$, which in the absence of self interaction moves along a
circular and equatorial geodesic around a Schwarzschild black hole of
mass $M$. For this model, the local SF was computed recently in
Refs.~\cite{burko-00,detweiler}. The scalar field $\Phi$
satisfies the wave equation $\square\Phi=-4\pi\rho$, where $\rho$ is the
scalar charge density. Solving this equation for the appropriate motion in
Schwarzschild spacetime (see Ref.~\cite{burko-00} for details), we obtain
the bare SF by  ${^{\rm bare}f}_{\mu}^{\rm
SF}=q\partial_{\mu}\Phi$. The bare force ${^{\rm bare}f}_{\mu}^{\rm SF}$
can be regularized using mode-sum regularization, a procedure which yields
the physical, finite part of the SF, $f_{\mu}^{\rm SF}$. 

We next choose $\epsilon=0.1$, and integrate to find the orbit. We
started from $r_{\rm initial}=30M$ and integrated inward toward the
innermost stable circular orbit (ISCO) at $r=6M$. (We do not consider 
the correction to the location of the ISCO, which is irrelevant to the
question of interest here.) At
intermediate points along the orbit we evaluated $f_{\mu}^{\rm SF}$ and 
${f_{\mu}^{\rm SF}}'$ using a best fit to a smooth function. 
Our results are displayed in Fig.~\ref{fig1}. We find that the orbit to
$O(1)$ decays slower than the orbit to $O(\epsilon^{-1})$. This is indeed
expected, as $f_r^{(1)}$, which does not contribute at
$O(\epsilon^{-1})$, is a repulsive force, which slows down the decay of the
orbit. Note, that because our correction to the orbit is at $O(1)$, the
last few revolutions look qualitatively the same also for other choices
of $\epsilon$. 

\begin{figure} 
\input epsf 
\epsfxsize=8.5cm 
\centerline{\epsfbox{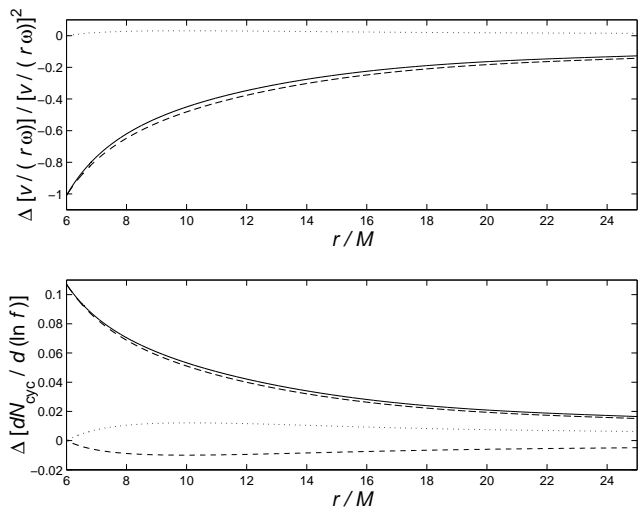}} 
\caption{The magnitude of the higher-order corrections. Upper
panel: $\Delta\left(\frac{V}{r\omega}\right)/\left(\frac{V}{r\omega}\right)^2$
as a function of $r$. The solid line is the full effect, the dotted
(dashed) line is the contribution of the term in
Eq.~(\ref{A}) proportional to $\sigma_{(1)}$ [$V_{(2)}$]. Lower
panel: $\Delta \,d{\cal N}_{\rm cyc}/\,d(\ln f)$ as a function of $r$. The
solid line is the full effect.
The dotted line is the contribution of the term in Eq.~(\ref{B})
proportional to $\sigma_{(1)}'$, and the lower (upper) dashed line the
contribution of the term proportional to $\sigma_{(1)}$ [$V_{(2)}$].}
\label{fig2}
\end{figure}

We next study the magnitude of the higher-order effect by considering Eqs. 
(\ref{B}) and (\ref{A}). Figure \ref{fig2} shows
$\Delta \,d{\cal N}_{\rm cyc}/\,d(\ln f)$ and 
$\Delta\left(\frac{V}{r\omega}\right)/\left(\frac{V}{r\omega}\right)^2$. We
first notice that both are dominated by the terms in the corresponding
equations proportional to $V_{(2)}$. In fact, at the ISCO the other terms
vanish, such that at the ISCO these quantities are fully described by the
terms proportional to $V_{(2)}$. (Strictly speaking, the ISCO may be
defined by the requirement that $\,d{\cal N}_{\rm cyc}/\,d(\ln f)$
vanishes. As noted above, we do not consider here the shift in the location 
of the ISCO, and our previous remarks relate, in fact, to $r=6M$, not the
ISCO proper. Note also that it is important to evaluate
$V_{(2)}$: the effect is controlled by that term at and near $r=6M$.) Fitting
our results, we find that 
$\Delta \,d{\cal N}_{\rm cyc}/\,d(\ln f)\sim D\, f^{\frac{8}{9}}$, where 
$D\approx 5.9$. Integrating from $r_{\rm start}=20M$ down to $r=6M$, we
find that the $O(\epsilon^{-1})$ template would slip by about a tenth of
a cycle compared with the data stream. 
This magnitude is independent of $\epsilon$. 
$V/(r\omega)$, however, is at $O(\epsilon)$. We thus conclude
that changes in amplitude will generally be unimportant for small values
of $\epsilon$. 

\begin{figure} 
\input epsf 
\epsfxsize=8.5cm 
\centerline{
\epsfbox{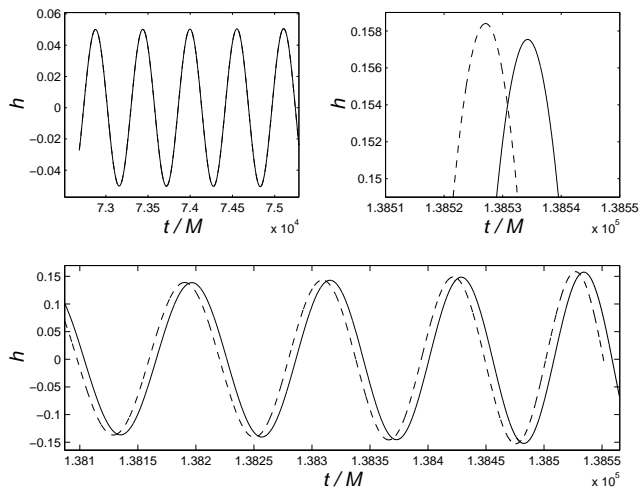}} 
\caption{The waveforms. In all panels the solid (dashed) 
line describes the waveform to $O(1)$ [$O(\epsilon^{-1})$]. The upper left
panel describes the waves from an early part of the orbit. 
The lower panel describes the last few cycles before the ISCO. The upper
right panel is a magnification of the last maximum.} 
\label{fig3}
\end{figure}

Finally, we present in Figure \ref{fig3} the waveforms to $O(1)$ and to
$O(\epsilon^{-1})$. We estimate the waveforms by means of the
usual ``restricted waveform'' approximation \cite{damour}, i.e.,
\begin{equation} h(t)=C\; v_{\omega}^2(t)\cos [\phi_{\rm wave} (t)]\, ,   
\label{waveform} \end{equation} where $v_{\omega}=(\,d\phi / \,dt)^{1/3}$,
and $C$ is an amplitude coefficient which depends on the distance of the
source of the waves. (For gravitational waves $\phi_{\rm  
wave}=2\phi$, and that for scalar field waves $\phi_{\rm wave}=\phi$.)  
Notice that our results above are independent of our adoption of the
restricted waveform. The two waveforms
are initially in phase. A phase difference, described by
Eq.~(\ref{B}) builds up slowly, and is clearly visible in the last few
cycles. In addition a difference in amplitude is also clearly seen. The
latter effect, however, is an artifact of the relatively large value of
$\epsilon$. This change in amplitude decreases linearly with
$\epsilon$. The phase shift, on the other hand, is independent of
$\epsilon$.

We emphasize that our conclusions are specific to scalar field RR. When
gravitational RR is considered, the magnitude of the higher-order effects 
could be different. For gravitational RR
we pick up an obvious factor of $2$ in  Eq.~(\ref{B}), such that one
may expect the higher-order effect to be even more important for 
gravitational RR. Determination of $f_{\alpha}^{\rm SF}$ (and its
gradients) for gravitational RR is required in order to quantify this
effect accurately.

An important issue is the possibility to use templates to 
$O(\epsilon^{-1})$, to cross correlate against the data stream.
One may hope, that templates to $O(\epsilon^{-1})$ for slightly different
parameters may be a good approximation, such that one does not have to
build templates to $O(1)$. We believe that such an
approach may indeed be useful for search purposes. This is in general
impossible to do to high accuracy when the cross correlation is done with
the full wave train: The waveform is a power series in $\epsilon$, in
which there are different functions of $r$ at each order. When one changes
$\epsilon$ to correct for the $O(1)$ effects in the waveform, the
$O(\epsilon^{-1})$ terms change correspondingly, such that the possibility
to fit the template to the data stream by changing $\epsilon$ appears to
us to be slim. Indeed, numerical experiments suggest to us that when the
parameter space is one dimensional, such techniques are not very
useful. [In the scalar field toy model we have, in fact, a two-dimensional
parameter space, which includes $q/\mu$ and $\mu/M$, which can be varied
independently. In the more realistic case of gravitational RR, however,
$q/\mu$, where $q$ is loosely a charge for the gravitational field of
$\mu$ (or an active mass), is determined uniquely as unity by the
Equivalence Principle. Motivated by the gravitational RR problem, we
considered an effective one-dimensional parameter space also for the
scalar field toy model, and varied only $q^2/(\mu M)$.] On the other hand,
when more complicated orbits are concerned (e.g., non-equatorial orbits,
when both the particle and the central black hole are spinning), the
parameter space will be significantly larger, such that such an approach  
may be quite successful. Because of the very slow increase in the phase
difference between the $O(\epsilon^{-1})$ template and the data stream, it
appears to us that data analysis techniques based on dividing the wave
train into many chunks, in which the change in phasing is negligible, may
be very successful for this problem. However, we emphasize that for
realization of the potentially high accuracy gravitational-wave astronomy, 
one would have to take higher-order corrections, such as those considered
here, into consideration.

I thank Steve Detweiler for communicating to me his results before their
publication. I thank the participants of the Fifth Capra Ranch Meeting on
Radiation Reaction, and in particular Warren Anderson, Scott Hughes, and
Amos Ori for comments. I am indebted to Richard Price for 
invaluable discussions and useful suggestions.  This research was 
supported by the National Science 
Foundation through grant No.~PHY-9734871, and in part through the Center
for Gravitational Wave Physics, under Cooperative Agreement PHY-0114375.

\end{document}